\documentclass[prb,onecolumn,12pt]{revtex4-1}
\usepackage{epsfig}
\usepackage{amssymb}
\usepackage{threeparttable}
\usepackage{graphicx}
\usepackage{natbib}
\usepackage{longtable}
\usepackage{txfonts}
\usepackage{color}

\begin{document}
\title{Electronic, optical, and thermodynamic properties of borophene from first-principle calculations}
\author{Bo Peng$^1$, Hao Zhang$^{1,\dag}$, Hezhu Shao$^{2,\ddag}$, Yuanfeng Xu$^1$, Rongjun Zhang$^1$ and Heyuan Zhu$^1$}
\affiliation{$^1$Shanghai Ultra-precision Optical Manufacturing Engineering Center, Department of Optical Science and Engineering, Fudan University, Shanghai 200433, China\\
$^2$Ningbo Institute of Materials Technology and Engineering, Chinese Academy of Sciences, Ningbo 315201, China}

\begin{abstract}
Borophene (two-dimensional boron sheet) is a new type of two-dimensional material, which was recently grown successfully on single crystal Ag substrates. In this paper, we investigate the electronic structure and bonding characteristics of borophene by first-principle calculations. The band structure of borophene shows highly anisotropic metallic behaviour. The obtained optical properties of borophene exhibit strong anisotropy as well. The combination of high optical transparency and high electrical conductivity in borophene makes it a promising candidate for future design of transparent conductors used in photovoltaics. Finally, the thermodynamic properties are investigated based on the phonon properties.
\end{abstract}

\maketitle

\section{Introduction}

Two-dimensional (2D) materials are one of the most active areas of nanomaterials research due to their potential for integration into next-generation electronic and energy conversion devices \cite{Ferrari2015,Novoselov2012,Klinovaja2013,xu2014a}. Graphene, the most widely studied 2D material, is a zero-gap semiconductor with linear dispersion near the Dirac points. As a result, the charge carriers in graphene behave like massless Dirac fermions \cite{Geim2009}. In addition to the extremely high carrier mobilities, graphene only absorbs 2.3\% of visible light \cite{Nair06062008,Eigler20092936}. Thus, graphene may be a viable candidate for applications as a transparent conductor. Compared to the traditional indium tin oxide (ITO), graphene has several advantages in terms of weight, robustness and flexibility \cite{Wassei201052}.

Recently, a new type of 2D material, borophene (2D boron sheet), has been grown successfully on single crystal Ag(111) substrates under ultrahigh-vacuum conditions, and attracted tremendous interest due to their extraordinary properties \cite{Mannix18122015}. As scanning tunneling spectroscopy measurements demonstrated, borophene shows anisotropic metallic behaviour. Furthermore, borophene is predicted to have extraordinary mechanical properties, which may rival graphene \cite{Mannix18122015}. In addition, borophene has its own advantage over graphene: due to the strongly anisotropic structure, the electronic and magnetic properties of borophene can be orientation controlled for flexible applications \cite{Meng2016}. Therefore, a comprehensive understanding of the electronic, bonding, and thermodynamic properties of borophene is needed for applications in future devices.

Inspired by the potential application of borophene, we perform first-principle calculations to study the structural and electronic properties of borophene. We also investigate the chemical bonding in detail. Finally, the vibrational and thermodynamic properties of borophene are analyzed using density functional perturbation theory (DFPT).

\section{Method and computational details}

The calculations are performed using the Vienna \textit{ab-initio} simulation package (VASP) based on density functional theory (DFT) \cite{Kresse1996,Kresse1996a}. The exchange-correlation energy is described by the generalized gradient approximation (GGA) using the Perdew-Burke-Ernzerhof (PBE) functional \cite{Perdew1996}. To correct the intrinsic band problem in DFT, hybrid functional methods based on Heyd-Scuseria-Ernzerhof (HSE06) method are adopted \cite{HSE1,HSE2,HSE3}. In the HSE06 method, a fraction of the exact screened Hartree-Fock (HF) exchange is incorporated into the PBE exchange using a mixing parameter $\alpha = 0.25$. The wave functions between the cores are expanded in plane waves with a kinetic energy cutoff of 500 eV. A 25$\times$15$\times$1 \textbf{k}-mesh is used during structural relaxation for the unit cell until the energy differences are converged within 10$^{-6}$ eV, with a Hellman-Feynman force convergence threshold of 10$^{-4}$ eV/\AA. For monolayer borophene, we used periodic boundary conditions along the three dimensions, and the vacuum space is around 15 \AA\ along the $z$ direction, which is enough to avoid the interaction between periodical images. Then the electronic structure, chemical bonding and optical properties are calculated. The partial occupancies are treated using the tetrahedron methodology with Bl\"ochl corrections. The harmonic interatomic force constants (IFCs) are obtained using DFPT within both supercell and linear response approach \cite{DFPT}. The phonon dispersion and thermodynamic properties are calculated from the harmonic IFCs using the PHONOPY code \cite{Togo2008,Togo2015}. A 7$\times$5$\times$1 supercell with 7$\times$5$\times$1 \textbf{k}-mesh is used to ensure the convergence.

\begin{figure}[h]
\centering
\includegraphics[width=\linewidth]{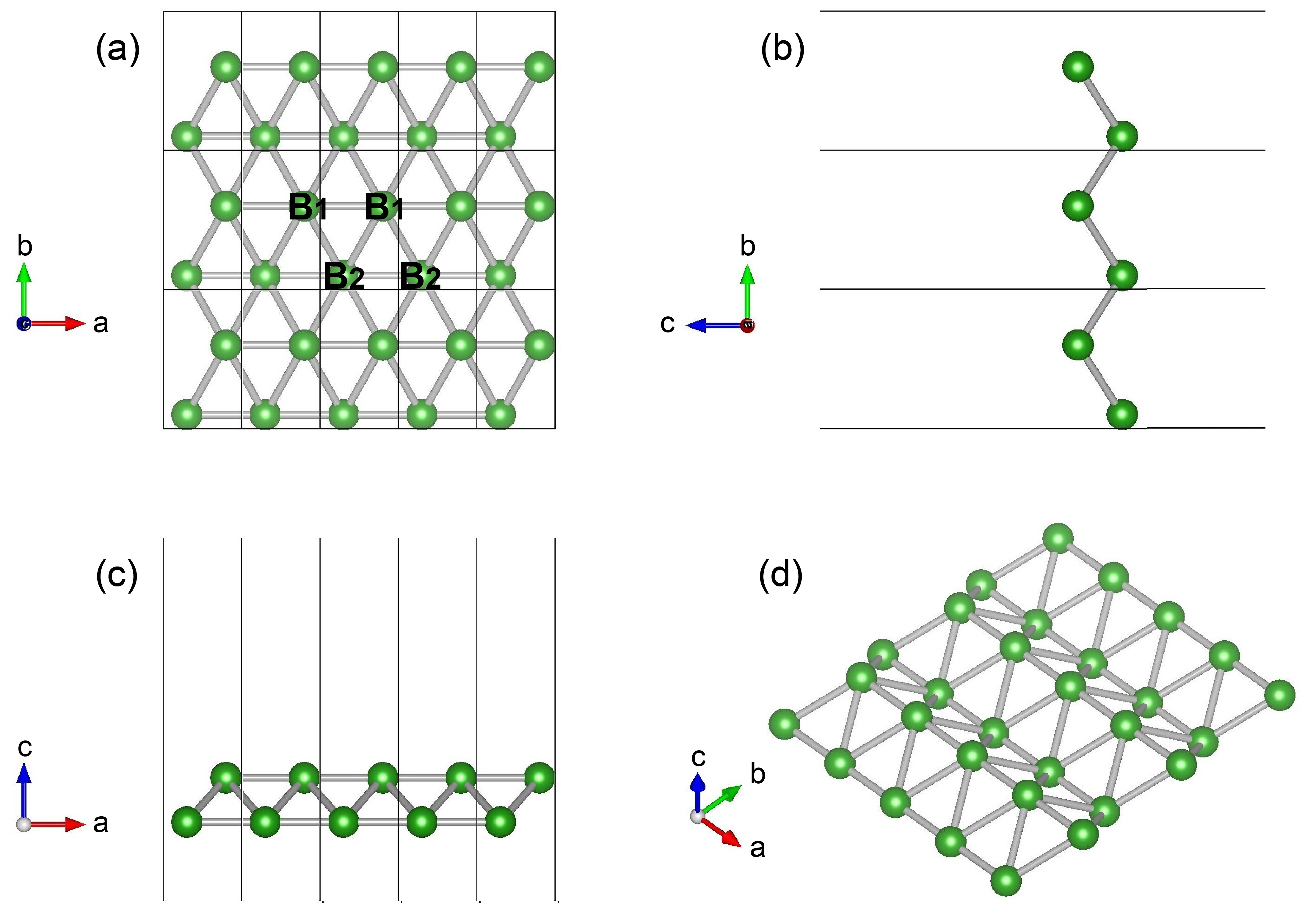}
\caption{(a) Top, (b) side, (c) front and (d) three-dimensional views of the atomic structure of borophene.}
\label{structure} 
\end{figure}

\section{Results and discussion}

\subsection{Structural and electronic properties of borophene}

Fig.~\ref{structure} shows the optimized structure of borophene. The optimized lattice constants are $a$=1.613 \AA\ and $b$=2.864 \AA.  Table~\ref{lattice} lists other typical structures for borophene as predicted in previous studies \cite{Zhou2014,Zhou2016,Carrete2016}. The predicted $a$ corresponds to 1/3 the $a$ observed in Ref. [9], while the predicted $b$ is in good agreement with experimental results \cite{Mannix18122015}. Although theoretical studies have proposed various structures for borophene \cite{Boustani1997,Tang2007,Lau2007,Liu2013a,Liu2013b,Zhou2014,Li2015,Yuan2015,Zhang2015c,Zhou2016,Carrete2016}, scanning tunneling microscopy measurements have shown that borophene has planar structure with anisotropic corrugation \cite{Mannix18122015}. There is no corrugations along the $a$ direction, while the buckling along the $b$ direction is observed. The predicted buckling height $h$ is 0.911 \AA. The bond length of B$_1$-B$_1$ and B$_2$-B$_2$ bonds along the $a$ direction is 1.613 \AA, and that of B$_1$-B$_2$ bonds is 1.879 \AA. In contrast to 2D honeycomb materials \cite{CastroNeto2009,Liu2011a,peng2016a,peng2016b}, borophene has a highly anisotropic crystal structure with space group $Pmmn$ \cite{Mannix18122015}.

\begin{table*}
\centering
\caption{Calculated lattice constants $a$ and $b$, buckling height $h$ of borophene. Other theoretical data and the experimental valus (from Ref. [9]) are also listed in parentheses for comparison.}
\begin{tabular}{cccc}
\hline
 Phase & $a$ (\AA) & $b$ (\AA) & $h$ (\AA) \\
\hline
 $Pmmn$ in this work & 1.613 & 2.864 & 0.911 \\
 $\alpha$ sheet \cite{Zhou2014} & 5.07 & 5.07 & 0 \\
 $Pmmn$ \cite{Zhou2014} & 4.52 & 3.26 & 0.69 \\
 $Pmmm$ \cite{Zhou2014} & 2.88 & 3.26 & 1.17 \\
 nonmagnetic \cite{Zhou2016} & 5.178 & 5.178 & 3.814 \\
 ferromagnetic \cite{Zhou2016} & 5.194 & 5.194 & 3.800 \\
 antiferromagnetic \cite{Zhou2016} & 5.189 & 5.189 & 3.836 \\
 $Pmmn$ \cite{Carrete2016} & 4.52 & 3.25 & 2.21 \\
 Experiment \cite{Mannix18122015} & 5.1$\pm$0.2 & 2.9$\pm$0.2 & - \\
\hline
\end{tabular}
\label{lattice}
\end{table*}

\begin{figure}[h]
\centering
\includegraphics[width=0.7\linewidth]{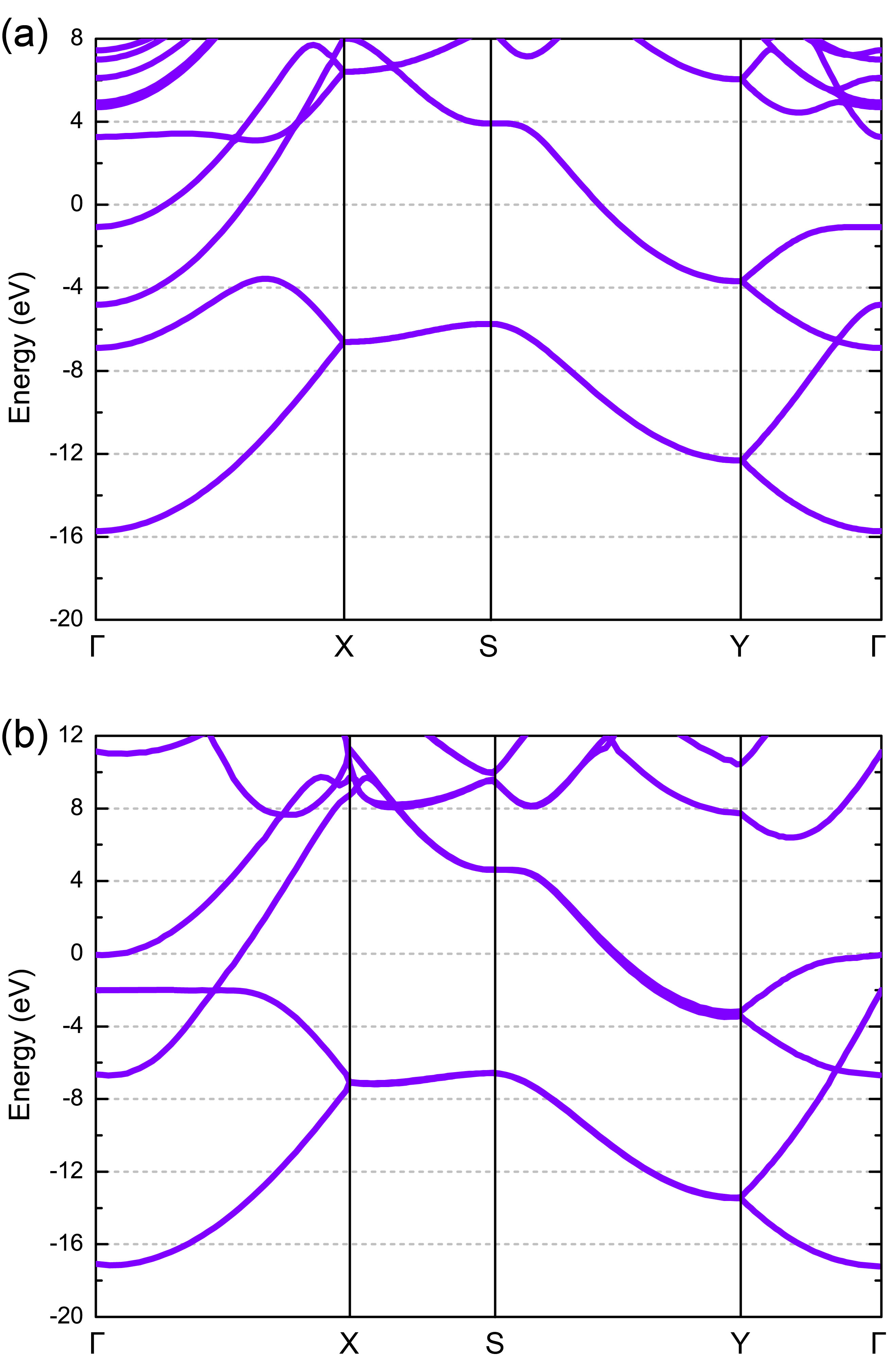}
\caption{Electronic band structure of borophene along $\Gamma$-X-S-Y-$\Gamma$ with (a) PBE and (b) HSE06 functional}.
\label{band} 
\end{figure}

The calculated band structure of borophene with the PBE functional along the high-symmetry directions of the Brillouin zone (BZ) is shown in Fig.~\ref{band}(a) with fixed Fermi level, which is in good agreement with previous theoretical work \cite{Mannix18122015}. The Fermi energy ($E_F$) is crossed by three different bands, one along S-Y direction, while the other two along $\Gamma$-X direction, indicating metallic behaviour along the directions parallel to the uncorrugated $a$ direction. However, the buckling along the $b$ direction opens a band gap of 9.66 eV and 4.34 eV along X-S and Y-$\Gamma$ directions, respectively. As a consequence, borophene behaves as a metal with strong anisotropy, and the electrical conductivity is confined along the uncorrugated $a$ direction.

The use of the HSE06 functional is important in obtaining the accurate electronic structure in borophene. Fig.~\ref{band}(b) shows the band structure of borophene with the HSE06 functional. Compared to the PBE functional, the HSE06 functional raises the conduction band minimum (CBM) about 0.72 eV and lowers the valence band maximum (VBM) 0.84 eV along X-S direction, increasing the band gap to 11.22 eV; while for Y-$\Gamma$ direction, the CBM is raised by 3.12 eV and the VBM is raised by 1.01 eV, increasing the band gap to 6.45 eV.

\begin{figure}
\centering
\includegraphics[width=\linewidth]{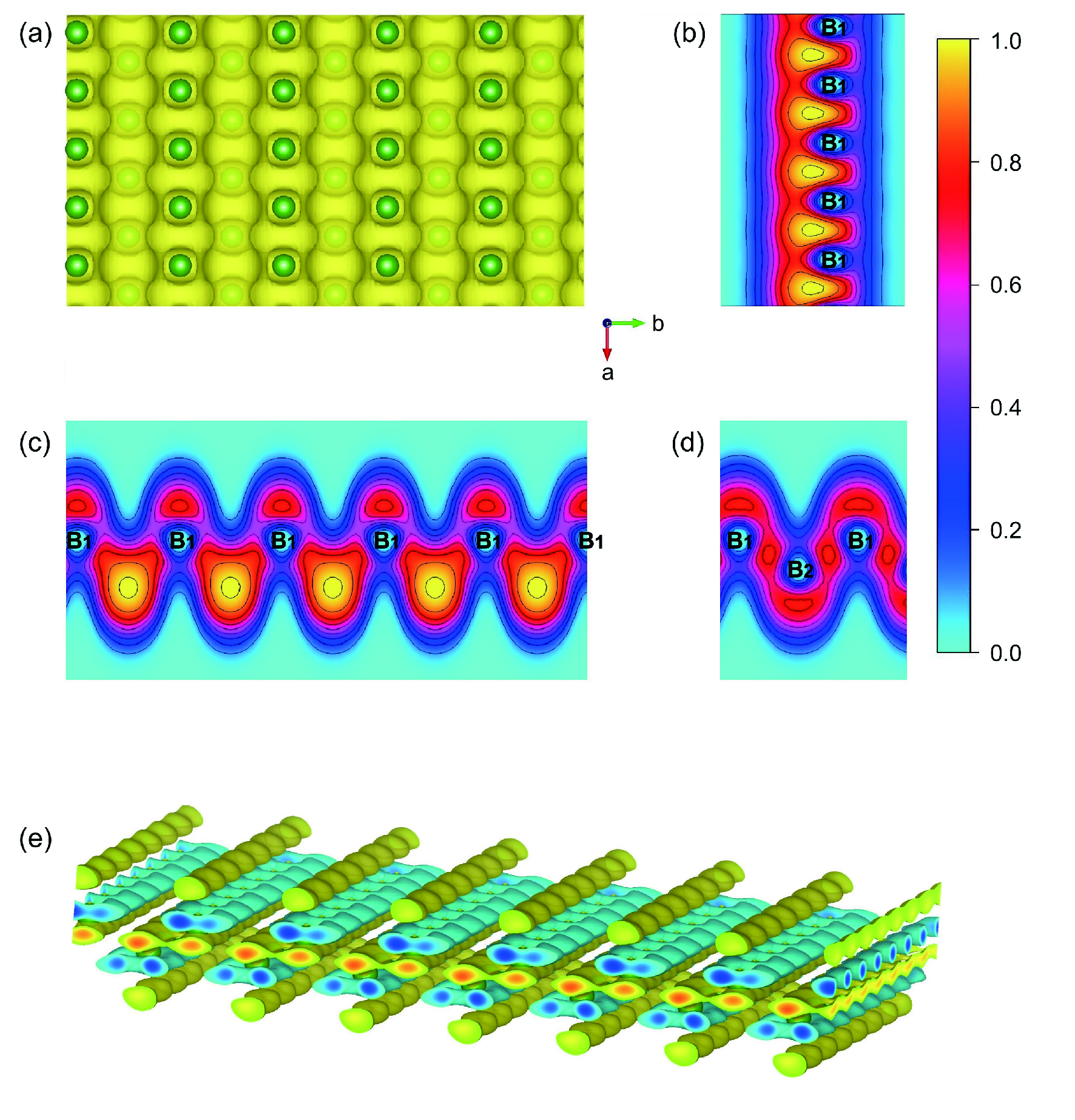}
\caption{(a) Top view of 3D ELF (isosurface=0.6) and 2D ELF profiles of borophene in the (b) [$0\bar{1}0$], (c) [$\bar{1}00$] and (d) [$\bar{1}\bar{1}0$] plane. (e) Spin density for borophene, where spin-up and -down components are represented by yellow and blue isosurfaces, respectively.}
\label{elf} 
\end{figure}

To understand the bonding characteristics, the electron localization function (ELF) \cite{Becke1990,Savin1992,Gatti2005,Chen2013} is calculated, as shown in Fig.~\ref{elf}. The ELF is a position dependent function with values that range from 0 to 1. ELF=1 corresponds to perfect localization and ELF=0.5 correponds to the electron-gas like pair probability. For the [$0\bar{1}0$] plane, the electrons are localized near B$_1$-B$_2$ bonds, indicating that the B$_1$-B$_2$ bond is strongly covalent. For the [$\bar{1}00$] plane, the electrons are accumulated on top of the upper plane and bottom of the lower plane, indicating the atomic orbitals of borophene contain more $sp^3$ hybridization, which is more stable than $sp^2$ hybridization. Similar to silicene and stanene \cite{Liu2011,Liu2011a,peng2016b}, the $sp^3$ hybridizations causes the 2D lattice of borophene along the $b$ direction to be buckled, which stabilizes the crystal structure.

The spin density is also demonstrated in Fig.~\ref{elf}(e), showing the magnetic structure. The spin-up and -down components are represented by yellow and blue isosurfaces, respectively. The up-spin states are well localized on top of the upper plane and bottom of the lower plane, while the down-spin states are accumulated near the B atom.

\begin{table}
\centering
\caption{Cohesive energies for borophene, graphene and silicene.}
\begin{tabular}{ccc}
\hline
  & Number of atoms &  Cohesive energy \\
  & (per unit cell) & (eV/atom) \\
\hline
Borophene & 2 & -5.99 \\
Graphene \cite{Quandt2008} & 2 & -10.13 \\
Silicene \cite{Drummond2012} & 2 & -4.57 \\
\hline
\end{tabular}
\label{cohesive}
\end{table}

In addition, we calculate the cohesive energy of borophene, as compared to graphene and silicene in Table~\ref{cohesive}. With lowest cohesive energy, the planar graphene is predicted to have the strongest interatomic bonding. The cohesive energy of the buckled silicene is higher than that of borophene with periodic vertical buckling, indicating a weaker Si-Si bond than B-B bond.

\subsection{Optical properties of borophene}

The optical properties of borophene is determined by the complex dielectric function, $i.e.$ $\epsilon(\omega)=\epsilon_1(\omega)+i\epsilon_2(\omega)$. For metals, the components of the dielectric tensor are given by a sum of interband and intraband contributions. Here we concern the visible region, and only interband transitions are taken into account, so there may be inaccuracy in dielectric function in the Drude region (low frequencies) \cite{Ehrenreich1962}. The imaginary part of dielectric tensor $\epsilon_2^{\alpha\beta}(\omega)$ is determined by a summation over empty band states using the equation \cite{Gajdos2006},
\begin{equation}
\epsilon_2^{\alpha\beta}(\omega) = \frac{2\pi e^2}{\Omega \epsilon_0} \sum_{k,v,c} \delta(E_k^c-E_k^v-\hbar \omega) \Bigg\vert\langle \Psi_k^c \big\vert \textbf{u}\cdot\textbf{r} \big\vert \Psi_k^v \rangle \Bigg\vert ^2,
\end{equation}
where $\epsilon_0$ is the vacuum dielectric constant, $\Omega$ is the volume, $v$ and $c$ represents the valence and conduction bands respectively, $\hbar\omega$ is the energy of the incident phonon, \textbf{u} is the vector defining the polarization of the incident electric field, \textbf{u}$\cdot$\textbf{r} is the momentum operator, and $\Psi_k^c$ and $\Psi_k^v$ are the wave functions of the conduction and valence band at the $k$ point, respectively. The real part of dielectric tensor $\epsilon_1^{\alpha\beta}(\omega)$ is obtained by the Kramers-Kronig relation,
\begin{equation}
\epsilon_1^{\alpha\beta}(\omega)=1+\frac{2}{\pi}P\int_0^{\infty} \frac{\epsilon_2^{\alpha\beta}(\omega ')\omega '}{\omega '^2-\omega^2+i\eta}d\omega ',
\end{equation}
where $P$ denotes the principle value. According to the dielectric function of borophene, the optical properties such as the energy loss spectrum $L(\omega)$, absorption coefficient $\alpha(\omega)$ and reflectivity $R(\omega)$ can be given by \cite{Saha2000,Luo2015}
\begin{equation}
L(\omega)=Im\Big(-\frac{1}{\epsilon(\omega)}\Big)=\frac{\epsilon_2(\omega)}{\epsilon_1^2(\omega)+\epsilon_2^2(\omega)},
\end{equation}
\begin{equation}
\alpha(\omega)=\frac{\sqrt{2}\omega}{c} \Big\lbrace \big[\epsilon_1^2(\omega)+\epsilon_2^2(\omega)\big]^{1/2}-\epsilon_1(\omega) \Big\rbrace ^{\frac{1}{2}},
\end{equation}
\begin{equation}
R(\omega)=\Bigg| \frac{\sqrt{\epsilon_1(\omega)+i\epsilon_2(\omega)}-1}{\sqrt{\epsilon_1(\omega)+i\epsilon_2(\omega)}+1} \Bigg| ^2.
\end{equation}

\begin{figure}
\centering
\includegraphics[width=0.8\linewidth]{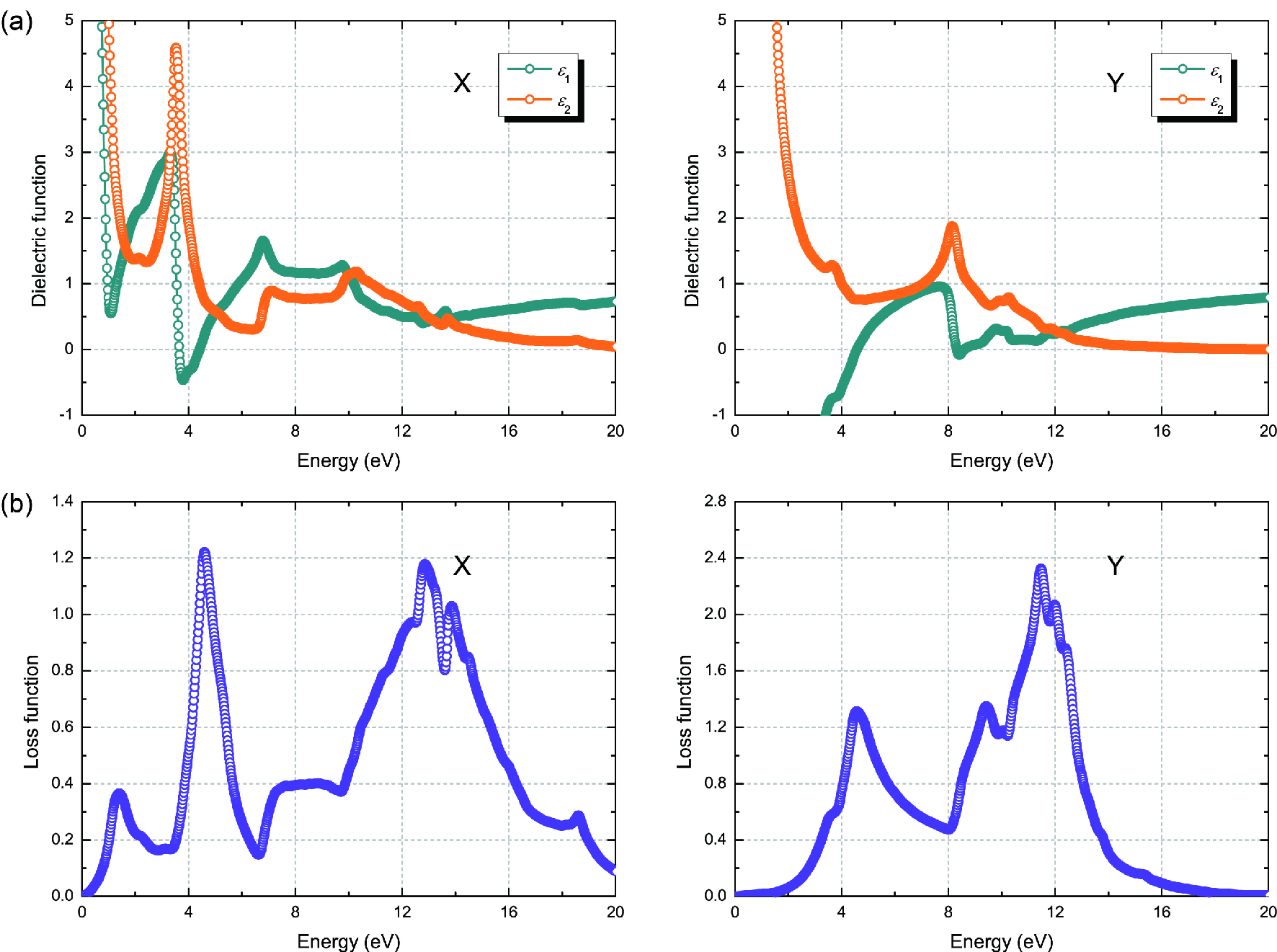}
\caption{(a) Dielectric function and (b) electron energy-loss function of borophene along the $a$ and $b$ directions.}
\label{dieletric} 
\end{figure}

The dielectric function and electron energy-loss function of borophene are calculated for incident radiations with
the electric field vector \textbf{E} polarized along the $a$ and $b$ directions in Fig.~\ref{dieletric}. Large in-plane anisotropy in optical properties is observed, which is attributed to the anisotropic crystal structure of borophene.

For both \textbf{E}//$a$ and \textbf{E}//$b$, the imaginary part of the dielectric function in Fig.~\ref{dieletric}(a) falls off rapidly at low frequencies near the free-electron region. For light polarized along X, $\epsilon_2(\omega)$ increases rapidly at 2.41 eV, and reaches the maximum at 3.52 eV. The peaks of $\epsilon_2(\omega)$ at 3.52 eV is probably due to the `parallel band' effect along $\Gamma$-X direction \cite{Fox2001}: when there is a band above $E_F$ that is approximately parallel to another band below $E_F$, the interband transitions from a large number of occupied $k$ states below $E_F$ occur at the same energy, which results in a strong peak. For light polarized along Y, the peaks of $\epsilon_2(\omega)$ at 8.14 eV is stronger than other peaks.

The energy-loss spectrum in Fig.~\ref{dieletric}(b) describes the energy loss of a fast electron traversing the material. For \textbf{E}//$a$, two prominent peaks are found at 4.59 eV and 12.85 eV respectively, which correspond to the free electrons plasmon peak. These represent the energy of collective excitations of the electronic charge density in the crystal. For \textbf{E}//$b$, the main peak is located at 4.55 eV and 11.47 eV.

\begin{figure}
\centering
\includegraphics[width=0.8\linewidth]{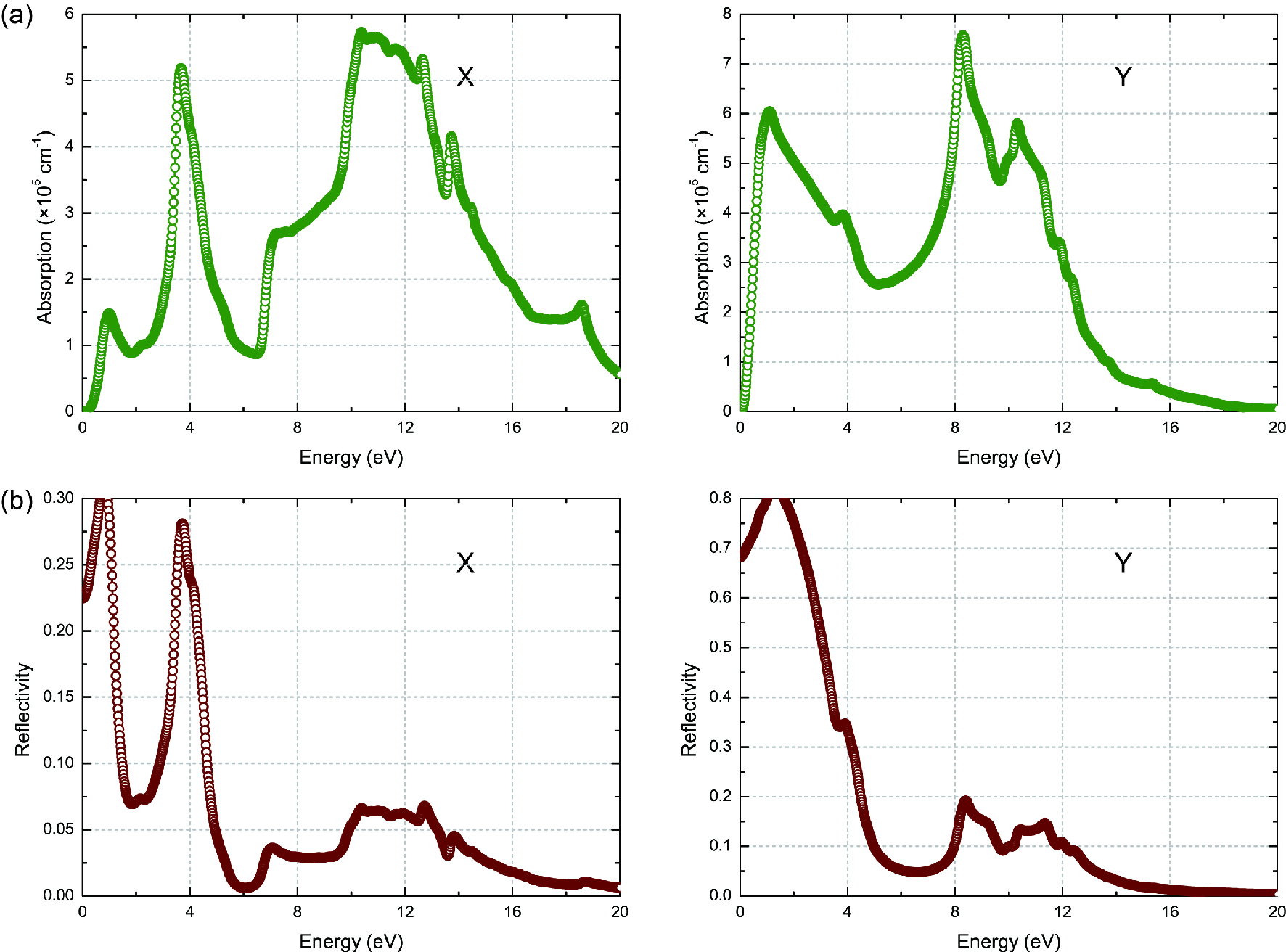}
\caption{(a) Absorption coefficient, and (c) reflectivity of borophene along the $a$ and $b$ directions.}
\label{absorption} 
\end{figure}

Fig.~\ref{absorption} shows the absorption coefficient and reflectivity. As shown in the absorption spectra in Fig.~\ref{absorption}(a), borophene has several absorption regions for \textbf{E}//$a$ around 3.65 eV and 10.36 eV. For \textbf{E}//$b$, three main absorption peaks are observed at about 1.09, 8.29 and 10.31 eV.

In the visible region, the reflectivity of borophene is lower than 30\% for \textbf{E}//$a$, while higher than 40\% for \textbf{E}//$b$, indicating that the optical properties of borophene can be orientation controlled. Such properties provide opportunities for applications in display technologies, photovoltaics and flexible electronics.

\begin{figure}
\centering
\includegraphics[width=\linewidth]{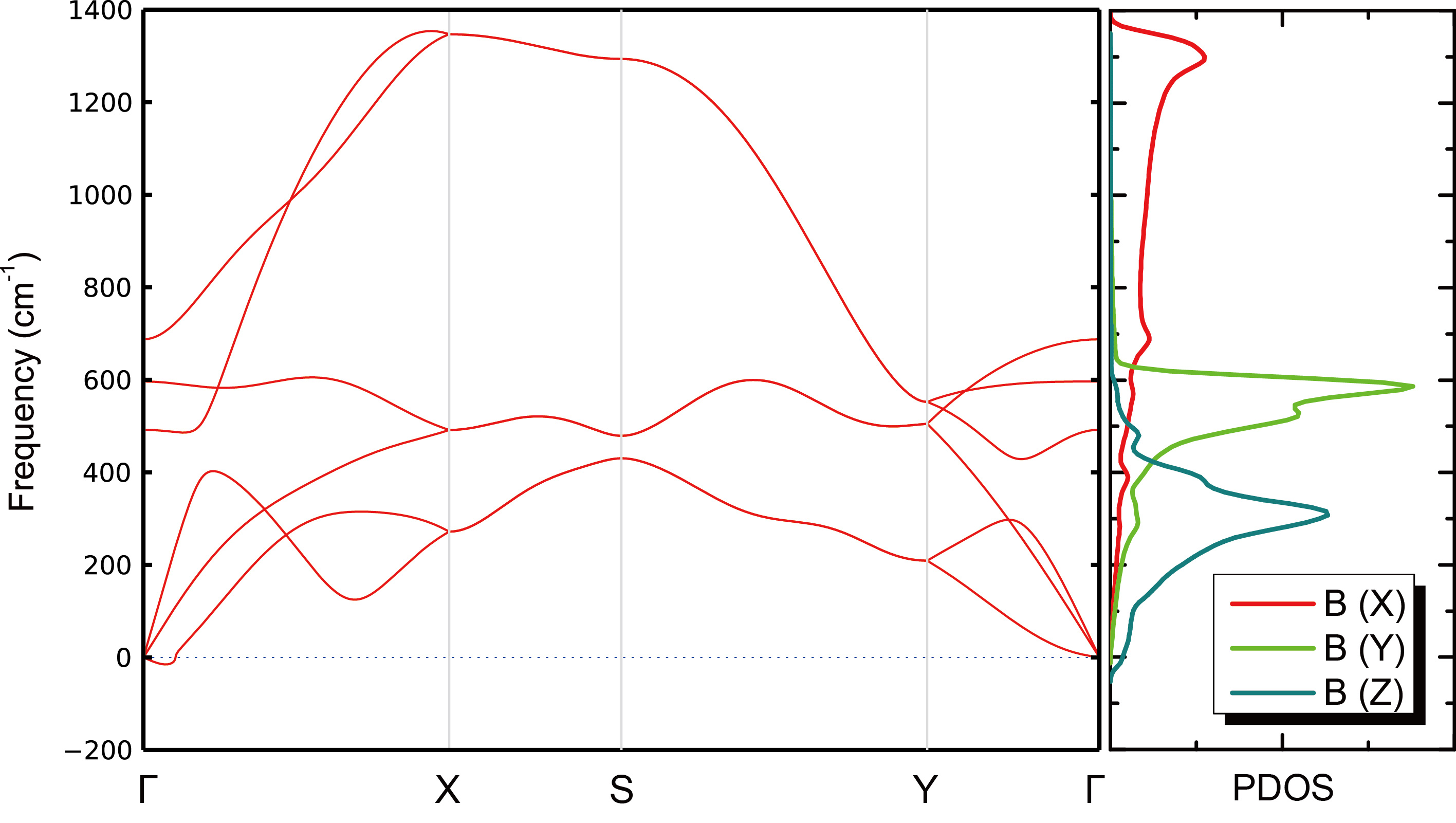}
\caption{Phonon spectrum and projected PDOS for borophene.}
\label{phonon} 
\end{figure}

\subsection{Lattice-dynamical properties of borophene}

Fig.~\ref{phonon} presents the phonon spectrum
along several high symmetry directions, together with the corresponding projected phonon density of states (PDOS). The primitive cell of borophene contains 2 atoms, corresponding to three acoustic and three optical phonon branches. Similar to other 2D hexagonal materials \cite{Nika2009,Qin2015,Liu2015,pb1,peng2016a}, the longitudinal acoustic (LA) and transverse acoustic (TA) branches are linear in the vicinity of the $\Gamma$ point, while the out-of-plane acoustic branch is quadratic along $\Gamma$-Y direction. However, the ZA branch has imaginary frequency along $\Gamma$-X direction. Negative frequency means that the restoring forces cannot be generated for the ZA branch \cite{Pang2016}. It indicates that the lattice exhibits instability for long-wavelength transverse thermal vibrations, which can explain the observed stripe formation along the $a$ direction in the experimental STM images \cite{Mannix18122015}. In fact, recent studies have suggested that the tensile strength of borophene is dictated by out-of-plane soft-mode phonon instability under biaxial tension or uniaxial tension along the $a$ direction \cite{Wang2016}, and free-standing borophene is instable even under high tensile stress \cite{Pang2016}.

The Debye temperature $\Theta_D$ can be calculated from the highest frequency of normal mode vibration (Debye frequency) $\omega_m$,
\begin{equation}
\Theta_D=\frac{\hbar \omega_m}{k_B},
\end{equation}
where $\hbar$ is the reduced Planck constant, and $k_B$ is the Boltzmann constant. The calculated Debye temperature for borophene is 863.86 K, which is higher than that of monolayer MoS$_2$ (262.3 K) \cite{peng2016a} and black phosphorene (500 K) \cite{Jain2015}, but lower than that of graphene (2,300 K) \cite{Efetov2010}. Concerning thermal vibrations, the Debye temperature is a measure of the temperature above which all modes begin to be excited \cite{Nakashima1992}, which will further affect the heat transport in borophene.

We also investigate the vibrational properties of borophene by calculating the PDOS for B(X), B(Y), and B(Z) vibrations as shown in Fig.~\ref{phonon}. The low-frequency acoustic phonon branches of borophene up to 450 cm$^{-1}$ are mainly from the B(Z) vibrations, while the high-frequency optical phonon branches are mainly from the B(X) vibrations. The B(Y) vibrations contribute significantly to the phonon DOS between 450 cm$^{-1}$ and 650 cm$^{-1}$.

\begin{figure}
\centering
\includegraphics[width=0.8\linewidth]{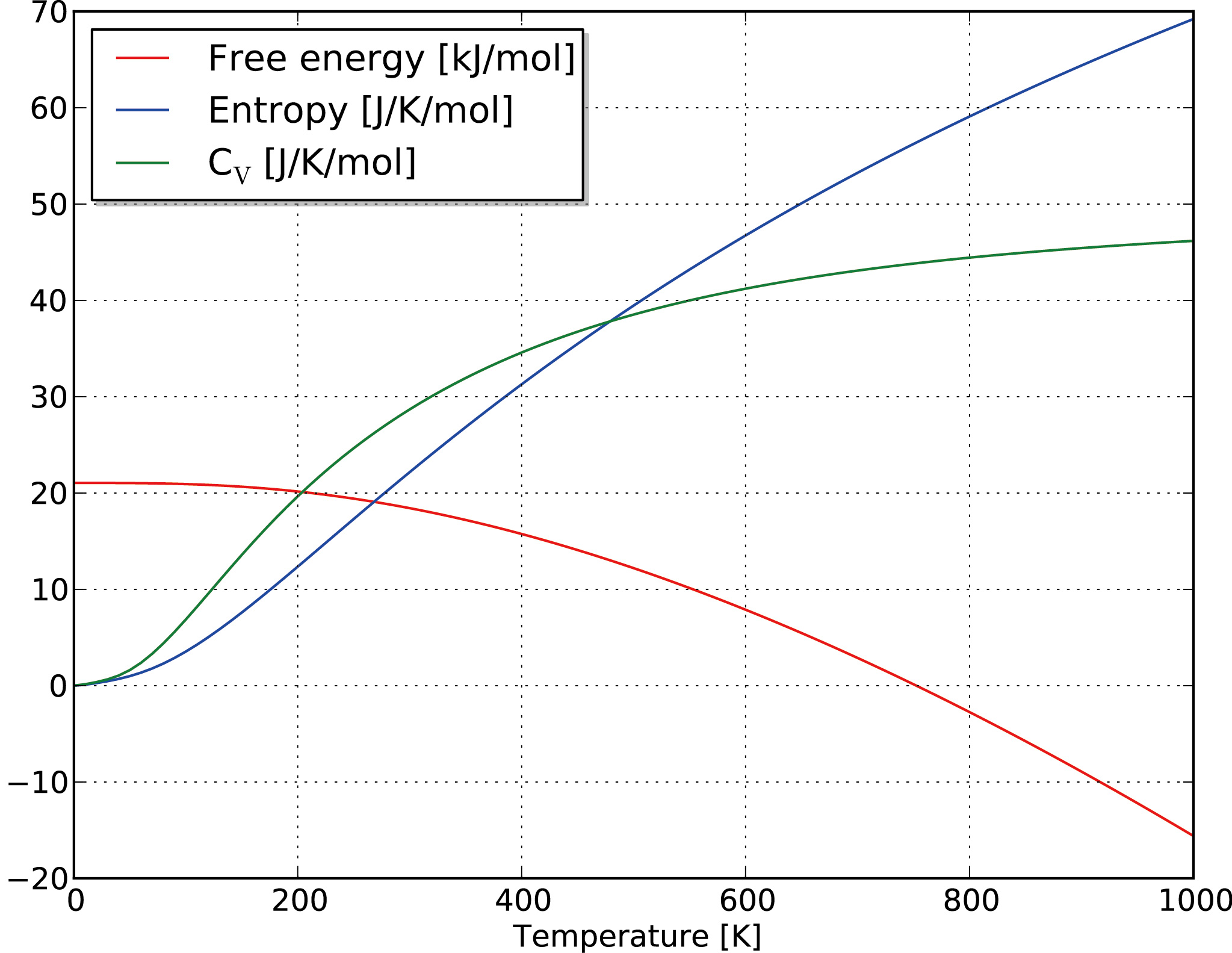}
\caption{Calculated Helmholtz free energy, entropy, and constant volume heat capacity for borophene.}
\label{thermal} 
\end{figure}

Furthermore, using phonon frequencies in the whole BZ, we calculate the thermodynamic properties such as Helmholtz free energy $F$, entropy $S$, and constant volume heat capacity $C_V$ \cite{Togo2008,Togo2015},
\begin{equation}
        F  =  -k_B T \ln Z 
                   =  \frac{1}{2} \sum\limits_{\textbf{q}j}\hbar\omega_{\textbf{q}j}+k_BT\sum\limits_{\textbf{q}j}\ln[1-\exp(-\hbar \omega_{\textbf{q}j}/k_BT)],
\end{equation}
\begin{equation}
        S =  \frac{\partial F}{\partial T}
                  =  \frac{1}{2T} \sum\limits_{\textbf{q}j}\hbar\omega_{\textbf{q}j}\coth [\hbar \omega_{\textbf{q}j}/2k_BT]-k_B\sum\limits_{\textbf{q}j}\ln[2\sinh(\hbar \omega_{\textbf{q}j}/2k_BT)],
\end{equation}
\begin{equation}
        C_V  =  \Big(\frac{\partial E}{\partial T}\Big)_V
                  =  \sum\limits_{\textbf{q}j}k_B \Big(\frac{\hbar\omega_{\textbf{q}j}}{k_BT} \Big)^2\frac{\exp(\hbar \omega_{\textbf{q}j}/k_BT)}{[\exp(\hbar \omega_{\textbf{q}j}/k_BT)-1]^2},
\end{equation}
where $Z$ is the partition functionq, \textbf{q} is the wave vector, and $j$ is the band index.

Fig.~\ref{thermal} shows the temperature dependence of the calculated Helmholtz free energy, entropy, and constant volume heat capacity for borophene. The free energy increases with increasing temperature, while the entropy decreases with increasing temperature. These two terms in Fig.~\ref{thermal} are zero at 0 K, which is in complete agreement with the third law of thermodynamics. The heat capacity approaches the Dulong-Petit classical limit (49.88 J/K/mol) at high temperatures.

\section{Conclusion}

In this work, we investigated the electronic structure, chemical bonding, optical and thermodynamic properties of borophene by first-principle calculations. In contrast to 2D honeycomb materials, borophene has a highly anisotropic crystal structure. The band structure predicts that borophene exhibits highly anisotropic metallic behaviour. We also discuss the bond characteristics of borophene. The interatomic bond strength in borophene is stronger than that in buckled silicene, but weaker than that in planar graphene.

The dielectric function, refractive index, conductivity, absorption coefficient, electron energy-loss spectrum and reflectivity are also calculated and discussed. Large optical anisotropy is observed in borophene due to the anisotropic crystal structure. There is no absorption in the visible region, and the reflectivity is very low. Due to the high optical transparency and electrical conductivity, as well as a variety of novel anisotropic properties, borophene can be used as transparent conductors for future applications in display technologies, photovoltaics and flexible electronics.

The phonon spectrum and PDOS are calculated using density functional perturbation theory. The negative ZA branch along $\Gamma$-X direction can explain the observed stripe formation along the $a$ direction in the experimental STM images. The Debye temperature of borophene is 863.86 K. The vibrational properties are investigated as well. Finally, the thermodynamic properties are determined using the phonon spectrum over the entire BZ.

\section*{Acknowledgement}
This work is supported by the National Natural Science Foundation of China under Grants No. 11374063 and 11404348, and the National Basic Research Program of China (973 Program) under Grants No. 2013CBA01505.




\providecommand*{\mcitethebibliography}{\thebibliography}
\csname @ifundefined\endcsname{endmcitethebibliography}
{\let\endmcitethebibliography\endthebibliography}{}

\end{document}